\documentclass[aps,showpacs,amssymb]{revtex4}
\usepackage[english]{babel}
\usepackage{bm}

\newcommand{\beq}{\begin{equation}}
\newcommand{\eeq}{\end{equation}}
\newcommand{\ben}{\begin{eqnarray}}
\newcommand{\een}{\end{eqnarray}}
\newcommand{\Asl}{\slash\kern-.65em A}
\newcommand{\psl}{\slash\kern-.55em p}
\newcommand{\delsl}{\slash\kern-.45em \partial}
\newcommand{\hsp}{\hspace{1.0cm}}

\newcommand{\noi}{\noindent}

\newcommand{\ce}{\frac{c}{e}}
\newcommand{\ec}{\frac{e}{c}}
\newcommand{\vp}{\mbox{\boldmath$p$}}
\newcommand{\vrr}{\mbox{\boldmath$x$}}
\newcommand{\vA}{\mbox{\boldmath$A$}}
\newcommand{\vf}{\mbox{\boldmath$f$}}
\newcommand{\M}{\mathcal{M}}

\newcommand{\vn}{\mbox{\boldmath$\nabla$}}
\newcommand{\la}{\mbox{\boldmath$\lambda$}}
\newcommand{\ppi}{\mbox{\boldmath$\pi$}}
\newcommand{\vB}{\mbox{\boldmath$B$}}
\newcommand{\vE}{\mbox{\boldmath$E$}}

\newcommand{\p}{\partial}

\begin{document}

\title{Coordinate noncommutativity in strong non-uniform magnetic fields}

\author{J. Frenkel}
\author{S. H. Pereira}
\affiliation{Instituto de F\'{\i}sica, Universidade de S\~ao Paulo, S\~ao Paulo, SP 05315-970, BRAZIL}

\bigskip

\begin{abstract}
\noi
Noncommuting spatial coordinates are studied in connection with
the motion of a
charged particle in a strong generic magnetic field. We derive a
relation involving the commutators of the coordinates, which
generalizes the one realized in a strong constant magnetic field. As
an application, we discuss the coordinate noncommutativity in a slowly
varying magnetic field.
\end{abstract}
 

\maketitle

\section{Introduction}

Noncommutativity of space coordinates has been much studied
from various points of view \cite{r1,r2,r3}. It arises naturally in string
theory, where it is related to the presence of a strong background
magnetic-like field. If this is constant, one obtains the more familiar case
where the coordinate noncommutativity $[x^i, x^j]$ is a constant antisymmetric quantity. However, if the background field
depends on the spatial coordinates, one would expect the coordinate noncommutativity to be a local function. This
possibility has been recently studied in the context of
noncommutative gauge field theories \cite{r4}.

On the other hand, coordinate noncommutativity may also arise in more
physical situations involving the motion of electric charges in a
strong external magnetic field \cite{r5,r6}. When a charged $(e)$ and massive
$(m)$ particle moves on a plane $(x,y)$ in the presence of a strong
constant magnetic field $B$ pointing along the $z$-axis, it has been
shown that the noncommutativity of space coordinates is of order
of the inverse of the magnetic field:
\beq
[x,y]=-i\hbar\frac{c}{eB}.\label{com}
\eeq
\noi
An interesting discussion of this behavior, which is related to the fact
that the large $B$ limit corresponds to small $m$, has been recently
given by Jackiw \cite{r7}.

Motivated by the above observations, we study in this note the motion of a
charged particle in a strong non-uniform magnetic
field $\vB(\vrr)$. Then, we argue that the relation (\ref{com}) can be
generalized to the rotationally symmetric form:
\beq
[x^i, x^j]=-i\hbar \ce \epsilon^{ijk}\frac{B_k(\vrr)}{B^2(\vrr)}\, ,
\hsp (i,j,k=1,2,3)\label{spc}
\eeq
\noi
which shows that the coordinate noncommutativity is
in this case a local function. 

This result for the coordinate noncommutativity in non-uniform
magnetic fields is derived in section 2. As an application, we study
in section 3 the behavior of noncommuting coordinates in a slowly varying magnetic field which is present,
for example, in a magnetic mirror.

\section{Noncommuting coordinates}

In order to derive the relation (\ref{spc}), we
consider the equation of motion of a charged particle in a static external
magnetic field:
\beq
m\mbox{\boldmath$\ddot{x}$}=\frac{e}{c}\mbox{\boldmath$\dot{x}$}\times
\vB(\vrr) +
\vf(\vrr),\label{fl}
\eeq
\noi
where $\vf(\vrr)$ represents
additional static forces which may be
derived from a potencial $V$: $\vf=-\mbox{\boldmath$\nabla$}$$V$. In the
presence of a strong magnetic field, the Lorentz force term can
dominate the kinetic term $m\mbox{\boldmath$\ddot{x}$}$, which therefore may be
dropped in first approximation. The resulting equation, however, cannot
determine all components of the velocity $\mbox{\boldmath$\dot{x}$}$, since
the projection of $\mbox{\boldmath$\dot{x}$}$ along $\vB$ is not specified in
(\ref{fl}). This is reflected in the equation:
\beq
(\mbox{\boldmath$\dot{x}$} \times\vB)_k + \frac{c}{e}f_k =
\epsilon_{kij}\dot{x}^iB^j + \frac{c}{e}f_k = 0, \label{fll}
\eeq
\noi
in that the antisymmetric matrix $(\epsilon_k)_{ij}$ does not
have an inverse. Multiplying (\ref{fll}) by $B^k$, we obtain the
consistency condition:
\beq
B^k f_k = \vB\cdot \vf = 0\,.\label{con}
\eeq
\noi
This relation ensures that the net force in the direction of $\vB$
vanishes, which represents a condition necessary to obtain, in the
limit $m
\to 0$, a consistent set of equations of motion. In fact, since the Lorentz force is orthogonal to the magnetic field,
this condition allows us to set the projection of
$m\mbox{\boldmath$\ddot{x}$}$ along $\vB$ equal to zero. The configuration described by equation (\ref{con}) may
be achieved provided the magnetic field is perpendicular to some
two-dimensional manifold $\M$. Then, if we take the potential $V$ to be a
function definied on $\M$, $\vf=-\mbox{\boldmath$\nabla$}$$V$ will be
tangencial to this manifold, so that the condition (\ref{con}) can be
satisfied.

One can see in a simple way that the form (\ref{spc}) for the
coordinate noncommutativity is consistent with the equation of motion
(\ref{fll}). To this end, let us consider the reduced Hamiltonian:
\beq
H_0=V(\vrr)\label{ham}
\eeq
\noi
which is obtained in the limit $m\to 0$, by setting the kinematical
momentum $m\mbox{\boldmath$\dot{x}$}$ equal to zero. Then, taking the Poisson bracket of $x^i$ with
$H_0$ and using the relation:
\beq
\dot{x}^i=\{x^i, H_0\}= f_j\{x^j, x^i\} \label{pbr}
\eeq
\noi
one can verify that the equation of motion (\ref{fll}) is satisfied
when the brackets which describe noncommuting coordinates are given by
the relation (\ref{spc}).

We shall now give a canonical derivation of
noncommutativity in the limit $m \to 0$, which is based on the
Hamiltonian:
\beq
H=\frac{\ppi^2}{2m} + V(\vrr)=\frac{1}{2m}(\vp-\ec \vA)^2 + V(\vrr) \label{hamm}
\eeq
\noi
where $\ppi$ is the kinematical momentum, $\vp$ is the canonical momentum and $\vA$ denotes the vector
potencial in the Maxwell theory. In
order to be able to set $m=0$ in (\ref{hamm}), we must impose $\ppi=0$
as a constraint. This can be implemented using Dirac's method for
dealing with constrained systems \cite{r8,r9} (for an alternative approach,
see reference \cite{r10}). Using this method, we
consider the constraints:
\beq
\pi^i= p^i - \ec A^i \approx 0 \hspace{2cm}(i=1,2,3) \label{constr}
\eeq
\noi
and evaluate their time evolution using the relation:
\beq
\dot{\pi}^i = \{\pi^i, H + \lambda_j \pi^j\}=\{\pi^i, V +
\lambda_j \pi^j\}=0 \label{constrr}
\eeq
\noi
where $\lambda_j$ represent the Lagrange multipliers in the
constrained theory. Using the canonical Poisson brackets, together
with the relation:
\beq
\{\pi^i, \pi^j\}=\ec (\p^i A^j - \p^j A^i)= \ec \epsilon^{ijk}B_k\, ,\label{phph}
\eeq
\noi
we obtain from (\ref{constrr}) the following set of equations involving these
multipliers:
\beq
\epsilon_{ijk}\lambda^iB^j - \ce \frac{\p V}{\p x^k} =
(\la \times \vB)_k + \ce f_k =0\,. \label{mult}
\eeq
\noi
This has the same structure as the one of the equation (\ref{fll}),
so that we may apply similar considerations as before. Namely, although this system leads to a consistent relation among the Lagrangian multipliers which implies the
condition (\ref{con}), it cannot determine all the $\lambda^i$ since
the projection of $\la$ along $\vB$ is not specified. One
can check this in more detail by writing $\la$ in terms of a linear
combination, with arbitrary coefficients, of the orthogonal vectors $\vB$,
$\vf$ and $\vB \times \vf$. Then, from equation (\ref{mult}) it
follows that $\la$ must actually have the form:
\beq
\la = \alpha \vB - \frac{c}{e} \frac{\vB \times \vf}{B^2}, \label{lam}
\eeq
\noi
so that the coefficient $\alpha$ remains undetermined. Using this
result, the total Hamiltonian in equation (\ref{constrr}) can be
written in the form:
\beq
H_t = V + \alpha \vB \cdot \ppi + \frac{c}{e}\frac{(\vf \times
\vB)}{B^2}\cdot\ppi  \label{ht}
\eeq
\noi
We note here that $c(\vf \times \vB)/eB^2$ represents the drift velocity
of the particle due to the force $\vf$. A well-known example of the
particle drift is the $\vE \times \vB$ drift which arises in a static
electric field $\vE$.

Since the coefficient $\alpha$ in the Hamiltonian (\ref{ht}) is
arbitrary, one may expect that:
\beq
\phi = \vB \cdot \ppi
\eeq
\noi
would be a first class constraint \cite{r8}, which commutes with all
constraints $\pi^i$. This is indeed the case, as one can easily check
with the help of equation (\ref{phph}).

Consequently, out of the three constraints $\pi^i$, we will be left
over with just two second-class constraints, which do not commute. We
may take these to be given by the following linear
combinations of the $\pi^i$:
\beq
\chi^1 = \vf \cdot \ppi \,\,\,; \hspace{1cm} \chi^2= (\vB \times
\vf)\cdot \ppi \label{qc}
\eeq
\noi
There is no loss of generality by this choice, since the vectors
$\vB$, $\vf$ and $\vB \times \vf$ are linearly
independent.

To proceed with the canonical formalism, we now introduce the Dirac
brackets:
\beq
\{x^i, x^j\}_D = \{x^i, x^j\}-\{x^i, \chi^k\}C_{kl}\{\chi^l, x^j\}\,,
\label{qd}
\eeq
\noi
where the matrix $C_{kl}$ is defined by:
\beq
C_{kl}\{ \chi^l, \chi^i\}= \delta^i_k\, . \label{qe}
\eeq
\noi
From equations (\ref{phph}) and (\ref{qc}) one can check, using the canonical
Poisson brackets, that:
\beq
\{x^i, \chi^1\}=f^i \,\,\,; \hsp \{x^i, \chi^2\}=\epsilon^{ijk}B_j f_k\,\,\, ;
\hsp \{\chi^1, \chi^2\} = \ec B^2 f^2\,. \label{qf}
\eeq
\noi
Then, with the help of equations (\ref{qe}) and (\ref{qf}), one finds
that the Dirac bracket (\ref{qd}) takes the form:
\beq
\{x^i, x^j\}_D=-\ce \epsilon^{ijk}\frac{B_k(\vrr)}{B^2(\vrr)}\, .\label{sc}
\eeq

One may pass over to the quantum theory, by taking the commutation
relations to correspond to $i\hbar$ times the Dirac bracket
relations. Then, from (\ref{sc}), one can verify the
result given in equation (\ref{spc}).

Examples of this type emerge on any 2D (co)adjoint orbit ${\cal M}$ (see
\cite{Pe, Ma, GP}), e.g. for a unit sphere $S^2$ with magnetic monopole in
its centre. The monopole magnetic field ${\vB}=B{\vrr}$, with ${\vrr}^2=1$,
gives the Dirac brackets (\ref{sc}) in the form $\{x^i,x^j\}_D=-\frac{c}{eB}\epsilon^
{ijk}x_k$. For discrete values of $B=\pm\ce\sqrt{s
(s+1)}$, $s$ half-integer, the quantization leads to the well-known
fuzzy sphere.

\section{Discussion}

The solution (\ref{spc}), which is symmetric under rotations in three
dimensions, describes noncommuting spatial coordinates in a generic
magnetic field. For consistency, such a noncommutative algebra must satisfy the
Jacobi identity:
\beq
\big[[x,y],z\big]+\big[[y,z],x\big]+\big[[z,x],y\big]=0\,. \label{jac}
\eeq
\noi
In order to show that this identity is satisfied, we note that the Jacobi identity
requires the condition:
\beq
\epsilon_{kij}[x^i, x^j]\hspace{0.1cm} \p_l [x^l, x^k] = 0\,. \label{da}
\eeq
\noi
Then, using the expression (\ref{spc}), we may write this condition in
the form:
\beq
\vB \cdot \vn \times \bigg( \frac{\vB}{B^2} \bigg)=0\,. \label{db}
\eeq
\noi
With the help of the Maxwell equation $\vn \times \vB = 0$
for the static external magnetic field, we can see that the above
equation is indeed satisfied.


As an application of the result (\ref{spc}), let us consider the case
of a slowly varying magnetic field in the $z$-direction. Such a field
occurs in a magnetic mirror \cite{r11} which confines the particle's motion
in the $z$-direction. It may be written in cylindrical coordinates in the
form:
\beq
\vB = -\frac{1}{2}\rho \frac{\p B_z(z)}{\p z}\mbox{\boldmath$\hat{e}_\rho$} +
B_z(z)\mbox{\boldmath$\hat{e}_z$}\,, \label{bfield}
\eeq
\noi
where $\rho B'_z << B_z$. Then, the solution (\ref{spc}) implies the
following relations among the noncommuting coordinates:
\beq
[x,y]=-i\hbar\ce \frac{B_z}{B^2} \hspace{0,3cm};\hspace{0,7cm}[y,z]=i\hbar\frac{c}{2e}
\frac{xB'_z}{B^2}\hspace{0,3cm}; \hspace{0,7cm}[z,x]=i\hbar\frac{c}{2e}
\frac{yB'_z}{B^2}\,.\label{spcc}
\eeq
\noi
We see that in this case the strongest coordinate noncommutativity
occurs in the $(x,y)$ plane and that the noncommutativity in the
$(x,z)$ and $(y,z)$ plane is weaker by a factor of order $\rho B'_z
/B_z <<1$.

As is well
known \cite{r12}, in the presence of a constant magnetic field along the
$z$-direction, the quantum energy levels of a charged particle are
given by:
\beq
E_{n,l}= \frac{eB_z}{2mc}\hbar(2n+|l|-l+1)+\frac{\hbar^2 k_z^2}{2m}\,, \label{energ}
\eeq
\noi
where $n=0,1,2 ...$ and $\hbar l$ gives the projection of the angular momentum on the
$z$-axis. The first term in (\ref{energ}) is associated with
the motion in the $(x,y)$ plane, and describes the Landau levels which
are infinitely degenerate. The
second term gives the translational energy of the particle
associated with its motion in the $z$-direction.

One can show that the relation (\ref{energ}) may also provide a good
approximation for the quantum energy levels of a charged particle in a
magnetic mirror, where $B_z$ is a slowly varying function of $z$. In
this case, one can see that as the particle drifts
along the $z$-axis, there will occur a gradual shift of the Landau
levels. This shift will be compensated by
a corresponding change in the translational energy of the particle,
so that its total energy remains conserved. We note that, since the separation
between the Landau levels is given by $\hbar e B_z/mc$, in a strong magnetic field only the
lowest Landau level is relevant. Furthermore, the large $B_z$ limit
is asymptotically equivalent to the limit $m\to0$. Hence, we may
interpret the coordinate noncommutativity (\ref{spcc}) as arising in
consequence of the fact that our system is constrained to lie in the
lowest Landau level.

\vspace{0,4cm}

\noi
We wish to thank A. Das and J. C. Taylor for reading the
manuscript and the referee for helpful comments. This work was supported by CAPES, CNPq and FAPESP, Brazil.

\end{document}